\documentclass[aps,prl,twocolumn,showpacs,preprintnumbers,amsmath,superscriptaddress,amssymb,floats,nofootinbib]{revtex4}

\usepackage{graphicx}
\usepackage{dcolumn}
\begin{document}

\def\Journal#1#2#3#4{{#1} {\bf #2}, #3 (#4)}
\def\NCA{\rm Nuovo Cimento}
\def\NIM{\rm Nucl. Instrum. Methods}
\def\NIMA{{\rm Nucl. Instrum. Methods} A}
\def\NPB{{\rm Nucl. Phys.} B}
\def\PLB{{\rm Phys. Lett.}  B}
\def\PRL{\rm Phys. Rev. Lett.}
\def\PRD{{\rm Phys. Rev.} D}
\def\PRC{{\rm Phys. Rev.} C}
\def\ZPC{{\rm Z. Phys.} C}
\def\JPG{{\rm J. Phys.} G}
\def\st{\scriptstyle}
\def\sst{\scriptscriptstyle}
\def\mco{\multicolumn}
\def\epp{\epsilon^{\prime}}
\def\vep{\varepsilon}
\def\ra{\rightarrow}
\def\ppg{\pi^+\pi^-\gamma}
\def\vp{{\bf p}}
\def\ko{K^0}
\def\kb{\bar{K^0}}
\def\al{\alpha}
\def\ab{\bar{\alpha}}
\def\be{\begin{equation}}
\def\ee{\end{equation}}
\def\bea{\begin{eqnarray}}
\def\eea{\end{eqnarray}}
\def\CPbar{\hbox{{\rm CP}\hskip-1.80em{/}}}
\title{\large \bf Nuclear Transparency with the
$\gamma~$n$~\rightarrow~\pi^{-}$~p Process in $^4$He.}
\author{D.~Dutta}
\affiliation{Duke University, Durham, NC 27708, USA}
\affiliation{Massachusetts Institute of Technology, Cambridge, MA 02139, USA}
\author{F.~Xiong}
\affiliation{Massachusetts Institute of Technology, Cambridge, MA 02139, USA}
\author{L.~Y.~Zhu} 
\affiliation{Massachusetts Institute of Technology, Cambridge, MA 02139, USA}
\author{J.~Arrington}
\affiliation{Argonne National Laboratory, Argonne, IL 60439, USA}
\author{T.~Averett}
\affiliation{College of William and Mary, Williamsburg, VA 23185, USA}
\affiliation{Thomas Jefferson National Accelerator Facility, Newport News, VA 23606, USA}
\author{E.~Beise}
\affiliation{University of Maryland, College Park, MD 20742, USA}
\author{J.~Calarco}
\affiliation{University of New Hampshire, Durham, NH 03824, USA}
\author{T.~Chang}
\affiliation{University of Illinois, Urbana, IL 61801, USA}
\author{J.~P.~Chen}
\affiliation{Thomas Jefferson National Accelerator Facility, Newport
News, VA 23606, USA}
\author{E.~Chudakov}
\affiliation{Thomas Jefferson National Accelerator Facility, Newport
News, VA 23606, USA}
\author{M.~Coman}
\affiliation{Florida International University, Miami, FL 33199, USA}
\author{B.~Clasie}
\affiliation{Massachusetts Institute of Technology, Cambridge, MA 02139, USA}
\author{C.~Crawford}
\affiliation{Massachusetts Institute of Technology, Cambridge, MA 02139, USA}
\author{S.~Dieterich}
\affiliation{Rutgers University, New Brunswick, NJ 08903, USA}
\author{F.~Dohrmann}
\thanks{Present address:FZ Rossendorf, Dresden Germany}
\affiliation{Argonne National Laboratory, Argonne, IL 60439, USA}
\author{K.~Fissum}
\affiliation{Lund University, S-221 00 Lund, Sweden}
\author{S.~Frullani}
\affiliation{INFN/Sezione Sanita, 00161 Roma, Italy}
\author{H.~Gao}
\affiliation{Duke University, Durham, NC 27708, USA}
\affiliation{Massachusetts Institute of Technology, Cambridge, MA 02139, USA}
\author{R.~Gilman}
\affiliation{Thomas Jefferson National Accelerator Facility, Newport News, VA 23606, USA}
\affiliation{Rutgers University, New Brunswick, NJ 08903, USA}
\author{C.~Glashausser}
\affiliation{Rutgers University, New Brunswick, NJ 08903, USA}
\author{J.~Gomez}
\affiliation{Thomas Jefferson National Accelerator Facility, Newport News, VA 23606, USA}
\author{K.~Hafidi}
\affiliation{Argonne National Laboratory, Argonne, IL 60439, USA}
\author{O.~Hansen}
\affiliation{Thomas Jefferson National Accelerator Facility, Newport News, VA 23606, USA}
\author{D.~W.~Higinbotham}
\affiliation{Massachusetts Institute of Technology, Cambridge, MA
02139, USA}
\author{R.~J.~Holt}
\affiliation{Argonne National Laboratory, Argonne, IL 60439, USA}
\author{C.~W.~de Jager}
\affiliation{Thomas Jefferson National Accelerator Facility, Newport News, VA 23606, USA}
\author{X.~Jiang}
\affiliation{Rutgers University, New Brunswick, NJ 08903, USA}
\author{E.~Kinney}
\affiliation{University of Colorado, Boulder, CO 80302, USA}
\author{K.~Kramer}
\affiliation{College of William and Mary, Williamsburg, VA 23185, USA}
\author{G.~Kumbartzki}
\affiliation{Rutgers University, New Brunswick, NJ 08903, USA}
\author{J.~LeRose}
\affiliation{Thomas Jefferson National Accelerator Facility, Newport News, VA 23606, USA}
\author{N.~Liyanage}
\affiliation{Thomas Jefferson National Accelerator Facility, Newport News, VA 23606, USA}
\author{D.~Mack}
\affiliation{Thomas Jefferson National Accelerator Facility, Newport News, VA 23606, USA}
\author{P.~Markowitz}
\affiliation{Florida International University, Miami, FL 33199, USA}
\author{K.~McCormick}
\affiliation{Rutgers University, New Brunswick, NJ 08903, USA}
\author{Z.-E.~Meziani}
\affiliation{Temple University, Philadelphia, PA 19122, USA}
\author{R.~Michaels}
\affiliation{Thomas Jefferson National Accelerator Facility, Newport News, VA 23606, USA}
\author{J.~Mitchell}
\affiliation{Thomas Jefferson National Accelerator Facility, Newport News, VA 23606, USA}
\author{S.~Nanda}
\affiliation{Thomas Jefferson National Accelerator Facility, Newport News, VA 23606, USA}
\author{D.~Potterveld}
\affiliation{Argonne National Laboratory, Argonne, IL 60439, USA}
\author{R.~Ransome}
\affiliation{Rutgers University, New Brunswick, NJ 08903, USA}
\author{P.~E.~~Reimer}
\affiliation{Argonne National Laboratory, Argonne, IL 60439, USA}
\author{B.~Reitz}
\affiliation{Thomas Jefferson National Accelerator Facility, Newport
News, VA 23606, USA}
\author{A.~Saha}
\affiliation{Thomas Jefferson National Accelerator Facility, Newport
News, VA 23606, USA}
\author{E.~C.~Schulte}
\affiliation{Argonne National Laboratory, Argonne, IL 60439, USA}
\affiliation{University of Illinois, Urbana, IL 61801, USA}
\author{J.~Seely}
\affiliation{Massachusetts Institute of Technology, Cambridge, MA
02139, USA}
\author{S.~\v{S}irca}
\affiliation{Massachusetts Institute of Technology, Cambridge, MA
02139, USA}
\author{S.~Strauch}
\affiliation{Rutgers University, New Brunswick, NJ 08903, USA}
\author{V.~Sulkosky}
\affiliation{College of William and Mary, Williamsburg, VA 23185, USA}
\author{B.~Vlahovic}
\affiliation{North Carolina Central University, Durham, NC 2770, USA}
\author{L.~B.~Weinstein}
\affiliation{Old Dominion University, Norfolk, VA 23529, USA}
\author{K.~Wijesooriya}
\affiliation{Argonne National Laboratory, Argonne, IL 60439, USA}
\author{C.~Williamson}
\affiliation{Massachusetts Institute of Technology, Cambridge, MA
02139, USA}
\author{B.~Wojtsekhowski}
\affiliation{Thomas Jefferson National Accelerator Facility, Newport
News, VA 23606, USA}
\author{H.~Xiang}
\affiliation{Massachusetts Institute of Technology, Cambridge, MA
02139, USA}
\author{W.~Xu}
\affiliation{Massachusetts Institute of Technology, Cambridge, MA
02139, USA}
\author{J.~Zeng}
\affiliation{University of Georgia, Athens, GA 30601, USA}
\author{X.~Zheng}
\affiliation{Massachusetts Institute of Technology, Cambridge, MA
02139, USA}
\collaboration{Jefferson Lab E94104 Collaboration}
\noaffiliation

\begin{abstract}
We have measured the nuclear transparency of the fundamental process 
$\gamma$~n $\rightarrow ~\pi^-$~p in $^4$He. These measurements 
were performed at Jefferson Lab in the photon energy range of 1.6~to~4.5~GeV and at $\theta^{\pi}_{cm}= 70^\circ$~and~$90^\circ$. These measurements are the 
first of their kind in the study of nuclear transparency in photoreactions. 
They also provide a benchmark test of Glauber calculations based on traditional models of nuclear physics. The transparency results suggest 
deviations from the traditional nuclear physics picture. The momentum 
transfer dependence of the measured nuclear transparency is 
consistent with Glauber calculations which include the quantum chromodynamics phenomenon of color transparency.  
\end{abstract}


\maketitle

Nuclear transparency is a very useful quantity for testing calculations based 
on traditional models of nuclear physics. It is defined as 
the ratio of the cross section per nucleon for a process on a bound nucleon in 
the nucleus to the cross section for the process on a free nucleon. It is also a typical quantity 
used in searches for deviations from the expectations of traditional 
nuclear physics, such as the phenomenon of Color Transparency (CT). CT refers 
to the vanishing 
of the final (and initial) state interactions of hadrons with the nuclear 
medium in exclusive processes at high momentum transfer~\cite{ct1}, and is a 
natural consequence of QCD. It is based on the idea that, at sufficiently high 
momentum transfer, the dominant amplitudes for exclusive reactions involve 
hadrons of reduced transverse size which can then  pass undisturbed through 
the nuclear medium. This is a novel QCD phenomenon which, if observed, would 
be a clear manifestation of hadrons fluctuating to a small size in the 
nucleus. Moreover, it also contradicts the traditional Glauber multiple 
scattering theory in the domain of its validity. Therefore, measurements of 
nuclear transparency have attracted a significant amount of effort during the 
last two decades. A clear signature for the onset of CT would involve a 
dramatic rise in the nuclear transparency as a function of  the 
momentum transfer involved in the process, i.e. a positive slope with 
respect to the momentum transfer.  

A number of searches for color transparency have been carried out in 
the last decade in experiments using the $A(p,2p)$ and $A(e,e'p)$ reactions 
and coherent and incoherent meson production from nuclei~\cite{ct6}~--~\cite{fermipi}. 
The $A(p,2p)$ nuclear transparency experiments carried 
out at Brookhaven~\cite{ct6} show a rise followed by a decrease
in the momentum transfer squared range of
$Q^2 \approx$  3 -- 10 (GeV/c)$^2$. This surprising behavior can be explained 
in terms of mechanisms other than color transparency~\cite{ct7,brodsky}. $A(e,e'p)$ experiments at SLAC~\cite{ne18} and more recently at JLab~\cite{ct5,hallc_ct}, have not found any evidence for an increase of the nuclear transparency 
up to a $Q^2$ value of 8.1 (GeV/c)$^2$. 
One would expect an earlier onset of CT for meson production than for proton 
scattering~\cite{blattel93}, as it is much more probable to produce a small
transverse size in a $q\bar{q}$ system than in a $qqq$ system. 
Experiments performed at Fermilab and DESY seem to support this 
idea~\cite{fermirho}~--~\cite{fermipi}. More recently, the HERMES 
collaboration~\cite{hermesrho} has reported a positive slope 
in the $Q^2$ dependence of nuclear transparency from coherent and 
incoherent $\rho^{0}$ production from nuclei at fixed coherence length.

In this letter, we report the first measurement of nuclear transparency of 
the $\gamma n \rightarrow \pi^- p$ process from $^4$He. There are several 
important advantages to the choice of the $^4$He nucleus and the 
$\gamma n \rightarrow \pi^- p$ process. 
Nucleon configurations obtained from the Monte Carlo method based 
on the exact nuclear ground state 
wavefunction are available for $^4$He~\cite{wiringa95}.  
These configurations along with the elementary hadron-nucleon 
cross-sections can be used to carry out precise calculations of the 
nuclear transparency~\cite{haiyan_t} in the framework of 
Glauber theory~\cite{glauber}. 
Therefore, precise measurement of nuclear transparency from $^4$He nuclei 
is a benchmark test of these traditional nuclear calculations and can be 
used to explore where the calculations start to breakdown. This could 
help identify the transition from the nucleon-meson degrees of freedom of the 
traditional nuclear physics to the quark-gluon degrees of freedom of QCD. 
Furthermore,
light nuclei such as $^4$He are predicted to be better for the search of 
CT phenomenon because of their relatively small nuclear 
sizes, which are smaller than the length scales over which the hadrons of 
reduced transverse size revert back to their equilibrium size.~\cite{ct3,ct4}.

The experiment was performed in Hall A~\cite{halla} at the Thomas Jefferson 
National Accelerator Facility (JLab). The continuous wave electron beam, with 
currents of approximately 30 $\mu$A  and energies ranging from 1.6 to 4.5 GeV, impinged 
on a $6\%$ copper radiator to generate an untagged bremsstrahlung photon beam. 
The combined photon and electron beam was then incident on a 15~cm target cell 
containing either helium or liquid deuterium. The two High Resolution Spectrometers (HRS) in Hall A, with a momentum resolution of better than $2\times 10^{-4}$ and a horizontal angular resolution of better than 2 mrad, were used to detect the outgoing pions and recoil protons in coincidence. The backgrounds 
from the electron beam and from the target cell walls were measured by taking data without the radiator inserted in the beam (only electron beam impinging on the production target) and also with an empty target cell inserted in the beam (both with and without the radiator inserted in the beam). 
Additional details on the experimental setup and the detectors used in this 
experiment can be found in Ref.~\cite{lingyan}.


\begin{figure}[htbp]
{\includegraphics*[width=9.0cm,height=9.0cm]{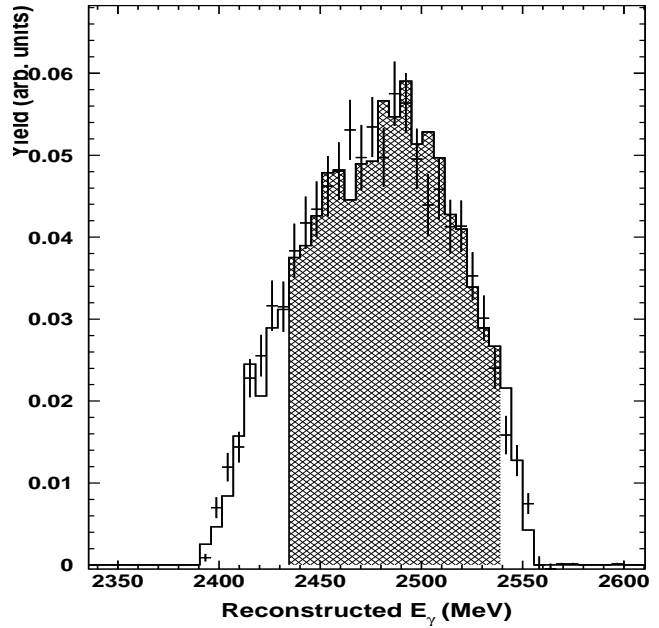}}
\caption[]{Reconstructed photon energy spectrum at 2.56 GeV and $\theta_{cm}= 90^\circ$. The curve is from the Monte Carlo simulation. The shaded area denotes the photon energy region used  to extract the experimental yield.} 
\label{compegamma}
\end{figure}

Based on two-body kinematics the incident photon energy is reconstructed for each event using the measured angles and momenta of the $\pi^-$ and $p$. In case of $^4$He we also assume that the residual nucleus is $^3$He. The resulting photon energy spectrum is a convolution of the bremsstrahlung distribution, the Fermi motion of the neutrons and the experimental acceptance. Cuts on trigger type, coincidence time, particle identification 
and acceptance were also applied while obtaining these spectra. A typical reconstructed photon energy spectrum is shown in Fig.~\ref{compegamma}. The experimental yield is obtained from these spectra by integrating 
over a 100~MeV window starting 25~MeV below the electron beam energy. 
This ensures that the contributions from multi-pion processes are negligible. The background yield from the electrons incident on a target were obtained 
by repeating the same procedure on data taken on that target without the 
radiator inserted in the beam. Similarly the background yields from the real 
photons and electrons incident on the target cell walls were obtained from 
the data taken for an empty target cell with and without the radiator inserted in the beam.

The background subtracted yield was then compared with the yield from a 
Monte Carlo simulation of the experiment with the same acceptance cuts. 
The Monte Carlo simulation was performed with the JLab Hall A Monte Carlo program, MCEEP~\cite{MCEEP}, which was adapted for photo-pion production 
experiments~\cite{lingyan}. The input angular distribution and cross-section 
used in MCEEP were obtained from a fit to the $\pi^+$ photo-production data at 4, 5 and 7.5 GeV~\cite{anderson}. The Fermi motion of the neutrons in the target nuclei was simulated using calculated momentum distributions (two-body breakup) and separation energy distributions of neutrons. For deuterium a calculated momentum distribution~\cite{momdist1} and fixed binding energy was used in the simulation. While for $^4$He, a calculated momentum distribution~\cite{momdist2} and an energy distribution based on the missing energy spectra measured in $^4$He(e,e'p) experiments for missing momentum $p_m$ = 100 $\pm$ 60 MeV/c~\cite{florizone}, was used. Additional details on the modifications to MCEEP for photo-pion production can be found in Ref.~\cite{lingyan}. 

The photon energy spectrum was reconstructed in the Monte Carlo simulation 
using the same method as used for the data, which includes the assumption in 
case of $^4$He that the residual nucleus is $^3$He. The quality of the simulation was studied by comparing the reconstructed angular and momentum distributions and the reconstructed photon energy spectrum with those obtained from the simulation. An example of the comparison of the reconstructed photon energy spectrum for a $^4$He target is shown in Fig.~\ref{compegamma}.

As per the definition of nuclear transparency one needs the cross section for $\gamma n \rightarrow \pi^- p$ reaction in $^4$He and in free space to extract transparency. However, since there are no free neutron targets we used a deuterium target and corrected for deuterium transparency. The transparency was 
extracted from the data and Monte Carlo yields from  $^4$He and $^2$H targets, using the relation:
\begin{equation}
T({\bf ^{4}He }) = \frac{\frac{{\mbox{Yield}}_{\mbox{Data}}({\bf ^{4}He})}{{\mbox{Yield}}_{\mbox{Monte Carlo}}({\bf ^{4}He})}}{\frac{{\mbox{Yield}}_{{\mbox{Data}}}({\bf ^{2}H})}{{\mbox{Yield}}_{\mbox{Monte Carlo}}({\bf ^{2}H})}}~~T({\bf ^{2}H })
\end{equation}

All data yields were corrected for computer dead time. A number of corrections such as pion decay, detector efficiencies and absorption in the 
spectrometer cancel when forming the ratio shown in Eq.~1. The ratio of the 
yields is corrected for the nuclear transparency of deuteron ($T_{^2H}$) 
which was obtained from the measured transparency of protons in $d(e,e'p)$ 
quasi-elastic scattering~\cite{hallc_ct} and a Glauber calculation~\cite{haiyan_t} of the transparency of $\pi^-$ in the deuteron. This correction was found 
to be on the order of 20\%. The point-to-point variation
of the transparency in the deuteron is negligible but there is a 3\% 
normalization systematic uncertainty associated with this correction. The
assumption that the residual nucleus is $^3$He, which is used in 
reconstructing the photon energy, introduces a normalization systematic 
uncertainty of $\approx$ 1.5\% and a point-to-point uncertainty of $<$0.5\%. 
This was determined from the fraction of Monte Carlo events which are 
generated from the tail of the input energy distribution above the two body 
breakup energy.
Another source of normalization systematic uncertainty is the neutron momentum and energy distribution used in the Monte Carlo simulation.
This was found to be 1\% for $^2$H and 2\% for $^4$He by using different calculated momentum distributions. The total 
normalization systematic uncertainty is 4.0\%.
 
In this procedure of extracting transparency using a super-ratio (Eq.~1), a 
number of systematic uncertainties such as charge, beam energy and 
bremsstrahlung photon yield cancel. This was checked 
rigorously by varying each of these quantities within their respective 
systematic uncertainties and then looking for the 
corresponding changes in the super-ratio. This test was also repeated on all 
the different cuts applied to the data, which were varied by 10--20\%. 
From these tests the point-to-point systematic uncertainty is estimated to 
be 2.7\% with most of the contribution coming from uncertainty in the 
target density due to local beam heating effects ($^2$H - 1\%, $^4$He - 1.5\%) 
and the energy loss calculation (1.4\%). Thus the total systematic 
uncertainty of the transparency measurement is 4.8\%.  

\begin{table}[htbp]
\caption[]{The extracted nuclear transparency for $\gamma n \rightarrow \pi^- p$ in $^4$He nucleus at  $\theta^{\pi}_{cm}$ = 70$^{\circ}$ and 90$^{\circ}$. There is an additional 4\% normalization systematic uncertainty and thus the total systematic uncertainty is 4.8\%. The $^2$H transparency used in the extraction is also shown.}
\begin{ruledtabular}
\begin{tabular}{|c|c|c|cc|c|}
E$_{\gamma}$& $|t|$ & T ($^4$He)& \multicolumn{2}{c}{Uncertainties}\vline& T($^2$H)\\ \hline 
 GeV & (GeV/c)$^2$ &   &  Stat & Pt.-Pt. Syst. &   \\ \hline
     &\multicolumn{5}{c}{$\theta^{\pi}_{cm}$ = 70$^{\circ}$}\vline \\ \hline
1.648 & 0.79& 0.583 & 0.008 & 0.015 & 0.815\\
2.486 & 1.28& 0.599 & 0.015 & 0.015 & 0.820\\
3.324 & 1.79& 0.628 & 0.013 & 0.016 & 0.815\\
4.157 & 2.31& 0.622 & 0.026 & 0.017 & 0.826\\\hline
     &\multicolumn{5}{c}{$\theta^{\pi}_{cm}$ = 90$^{\circ}$}\vline \\ \hline
1.648 & 1.20& 0.553 & 0.008 & 0.015 & 0.729\\
2.486 & 1.94& 0.559 & 0.012 & 0.015 & 0.812\\
3.324 & 2.73& 0.602 & 0.019 & 0.016 & 0.819\\
4.157 & 3.50& 0.614 & 0.026 & 0.017 & 0.827\\
\end{tabular}
\end{ruledtabular}
\label{trans_table1}
\end{table}

\begin{figure}[htbp]
{\includegraphics*[width=9.5cm,height=10cm]{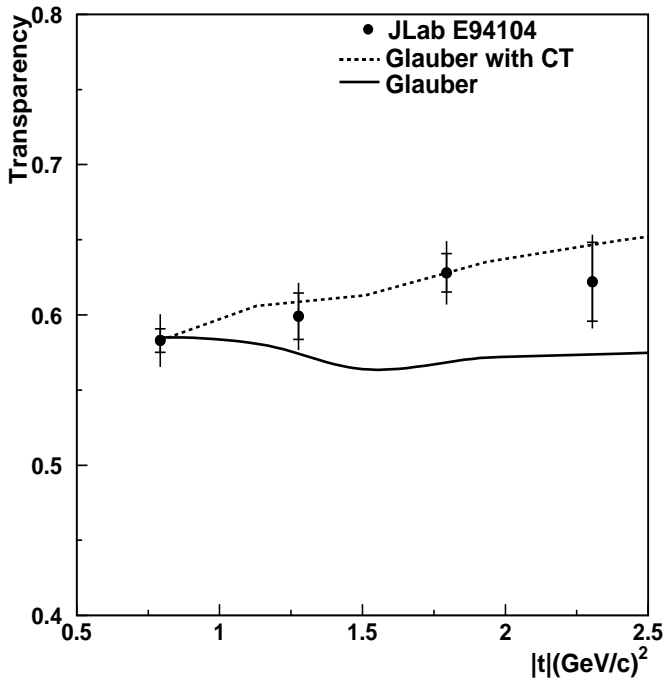}}
\caption[]{The nuclear transparency of $^4$He($\gamma$,p $\pi^-$) at   $\theta^{\pi}_{cm}=$70$^{\circ}$, as a function of momentum transfer square $|t|$. The inner error bars shown are statistical uncertainties only, while the outer error bars are statistical and point-to-point systematic uncertainties (2.7\%) added in quadrature. In addition there is a 4\% normalization/scale systematic uncertainty which leads to a total systematic uncertainty is 4.8\%.}
\label{trans70}
\end{figure}
\begin{figure}[htbp]
{\includegraphics*[width=9.5cm,height=10cm]{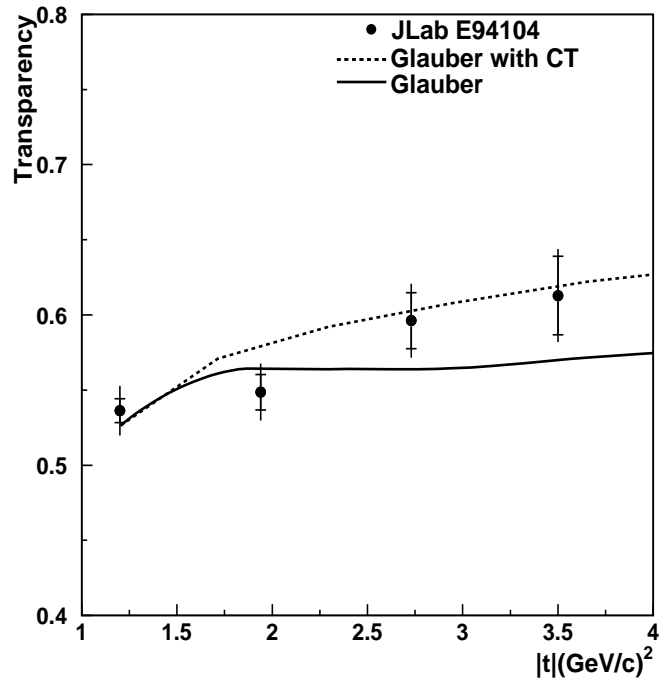}}
\caption[]{The nuclear transparency of $^4$He($\gamma$,p $\pi^-$) at  $\theta^{\pi}_{cm}=$90$^{\circ}$, as a function of momentum transfer square $|t|$. The inner error bars shown are statistical uncertainties only, while the outer error bars are statistical and point-to-point systematic uncertainties (2.7\%) added in quadrature. In addition there is a 4\% normalization/scale systematic uncertainty which leads to a total systematic uncertainty is 4.8\%.}
\label{trans90}
\end{figure}

The extracted nuclear transparency for the $^4$He target along with calculations is shown in Figs.~\ref{trans70} and ~\ref{trans90}; the results are also listed in Table~\ref{trans_table1}.
The Glauber calculation uses $^4$He configurations, which are snapshots of the positions of the nucleons in the nucleus, obtained from the variational wave 
function of Arriaga {\it et al.}~\cite{wiringa95}. 
These contain 
correlations generated by the Argonne $v_{14}$ and Urbana VIII models of the 
two body and three body nuclear forces respectively.  
The classical transparency was calculated from these configurations 
using the method described in Ref.~\cite{haiyan_t}. The hadron-nucleon total 
cross-sections were taken from Ref.~\cite{pdb}. The calculation which 
includes the CT effect was obtained by repeating the calculation mentioned 
above with the hadron-nucleon total cross-section modified according to 
the quantum diffusion model~\cite{ct3}. This procedure is also described in 
Ref.~\cite{haiyan_t} and was normalized to the Glauber calculation without CT at the lowest energy point (E$_{\gamma}$ = 1.648 GeV). 
There is $\approx$ 4-6\% uncertainty in both the Glauber and CT calculations arising from the uncertainty in the measured hadron-nucleon total cross-sections. 

A number of other CT calculations~\cite{strikma,nikola} have been performed for the $A(e,e'p)$ and $A(e,e'\pi)$ reactions. These 
different calculations generally predict 10~--~25\% effect  for the $^{12}$C$(e,e'p)$ reaction at a $Q^2$ = 10~(GeV/c)$^2$. Nevertheless, the positive slope of the transparency is very consistent among the different models.

In Fig.~2 and 3 the traditional nuclear physics calculation appears to 
deviate from the data at the higher energies. The absolute magnitude of the calculations with CT was normalized to the calculation without CT at the lowest 
energy point, however, it is the momentum transfer squared ($|t|$) dependence 
of the transparency which is of greater significance. The $|t|$ 
dependence is not affected by the normalization systematic uncertainties.  
The slopes of the measured transparency obtained from the three points which 
are above the resonance region (above $E_\gamma$ = 2.25 GeV ) are shown in
Table~\ref{slopetable}. 
\begin{table}[htbp]
\caption[]{The slope for the $|t|$ dependence of the extracted nuclear transparency obtained from the three points which are above the resonance region (above $\sqrt{s}$ = 2.25 GeV). The uncertainties are statistical and systematic respectively.}
\begin{ruledtabular}
\begin{tabular}{|c|c|c|c|}
 $\theta^{\pi}_{cm}$ & Measured slope & CT & Glauber\\ \hline
(deg) & (GeV/c)$^{-2}$ &  (GeV/c)$^{-2}$  &  (GeV/c)$^{-2}$\\ \hline
70 & 0.032~$\pm$~0.027~$\pm$~0.022 & 0.037 & 0.009\\
90 & 0.046~$\pm$~0.016~$\pm$~0.014 & 0.024 & 0.006\\ \hline 
\end{tabular}
\end{ruledtabular}
\label{slopetable}
\end{table}
These slopes are in good agreement, within experimental 
uncertainties, with the slopes predicted by the calculations with CT 
and they seem to deviate from the slopes predicted by the Glauber calculations at the 
$\approx$ 1~$\sigma$~(2~$\sigma$) level for  $\theta^{\pi}_{cm}=70^\circ$~($90^\circ$). The deviation from Glauber calculation is larger at $\theta^{\pi}_{cm}=90^\circ$, as expected for a CT-like effect, since it is at a higher pion  $|t|$. It is also interesting that the results 
are consistent with the rise expected for CT at the same photon energy 
at which  the onset of scaling behavior was observed in the cross-section for the $\gamma$~n~$\rightarrow~\pi^-$~p and 
the $\gamma$~p~$\rightarrow~\pi^+$~n processes~\cite{lingyan}.
Thus, these data suggest the onset of deviation from traditional calculations, 
but future experiments with significantly improved statistical and 
systematic precision are essential to put these results on a firmer basis.

In conclusion we have measured for the first time the nuclear transparency for the process $\gamma$~n $\rightarrow ~\pi^-$~p on a $^4$He target at  $\theta^{\pi}_{cm}=70^\circ$~and~$90^\circ$ 
in the photon energy range from 1.6 to 4.5 GeV. These measurements provide important tests for calculations based on the traditional model of nuclear 
physics and on Glauber theory. The measured
transparency show interesting momentum transfer squared dependence
which seem to deviate from the traditional nuclear physics predictions at the 
higher momentum transfers that suggests a CT-like behavior. A first indication of CT-like effect in this kind of reaction is interesting and calls for more data. Future experiments with better statistical 
and systematic precision in this energy range together with 
improved theoretical calculations are crucial for confirming these results.
       
We acknowledge the outstanding support of JLab Hall A technical staff and
Accelerator Division in accomplishing this experiment. We thank K.~Bailey
and T.~O'~Connor for their help with preparation of this experiment. 
We thank H.~Arenh$\rm \ddot{o}$vel and R.~B.~Wiringa for calculating the
momentum distribution of the neutron in the deuteron and $^4$He.
We thank Z.~Chai for providing the codes to apply R-function cut on
acceptance.
We also thank T.~W.~Donnelly, P.~Jain and G.~Miller for helpful discussions.
This work was supported in part by the U.~S.~Department of Energy,
DOE/EPSCoR,
the U.~S.~National Science Foundation,
the Ministero dell'Universit\`{a} e della Ricerca
Scientifica e Tecnologica (Murst),
the French Commissariat \`{a} l'\'{E}nergie Atomique,
Centre National de la
Recherche Scientifique (CNRS) and the Italian Istituto Nazionale di Fisica
Nucleare (INFN).
This work was supported by DOE contract DE-AC05-84ER40150
under which the Southeastern Universities Research Association
(SURA) operates the Thomas Jefferson National Accelerator Facility.


\end{document}